# Q-Factor Measurement of Nonlinear Superconducting Resonators

X. S. Rao[*], C. K. Ong and Y. P. Feng

Center for Superconducting and Magnetic Materials and Department of Physics,

National University of Singapore, Singapore 119260

**Abstract**

A novel method, which combined a multi-bandwidth measurement and an extrapolation procedure, is proposed for extracting the loaded Q-factor ($Q_L$) with improved accuracy from non-Lorentzian resonances of nonlinear superconducting resonators.

**Index Items**: microwave measurement, superconducting microwave devices, Q-factor, nonlinearity.

[*] Corresponding author. Tel: +65-8742615, Fax: +65-7776126, E-mail: scip7099@leonis.nus.edu.sg

## 1. Introduction

The nonlinear microwave surface impedance ($Z_S = R_S + jX_S$) of high temperature superconductors, i.e., its power dependence $Z_S(P)$, is of interest both for practical applications and for fundamental materials understanding. The study of $Z_S(P)$ often relies on the resonant techniques via measuring the Q-factor and the resonant frequency of a superconducting resonator as a function of microwave power $P$. There is a need for an accurate extraction of the Q-factors from the measured nonlinear resonance curves, which are generally non-Lorentzian as will be shown below, for quantifying $Z_S(P)$ and exploring its origins.

## 2. Method

In the case of linear response, the well-known frequency dependence of the transmission loss $T(f)=|S_{21}(f)|^2$ for a transmission mode resonator is [1]

$$T(f) = \frac{|T(f_0)|}{1 + [2Q_L(f-f_0)/f_0]^2} \quad (1)$$

where $T(f_0) = 4\beta_1\beta_2/(1+\beta_1+\beta_2)^2$, $\beta_1$ and $\beta_2$ are the coupling coefficients. The resonance curve of $T(f)$ has a well defined Lorentzian shape and the loaded Q-factor, $Q_L$, of the resonator can be easily obtained by measuring the 3-dB (half power) bandwidth, $\Delta f_{3dB}$, of the transmission curve and the resonant frequency $f_0$.

$$Q_L = f_0 / \Delta f_{3dB}. \quad (2)$$

When the resonator undergoes a nonlinear response, however, the resonance curve diverges from the Lorentzian shape and becomes asymmetric. This effect is illustrated in Fig.1 where the resonance curves of a microstrip resonator made from double-sided superconducting $YBa_2Cu_3O_{7-\delta}$ (YBCO) thin film are measured as the input power is increased in 5-dB steps from –18 to 12 dBm. Similar effects were also observed in the resonators made from NbN and Nb films [2]. At low input power, the resonance curves are symmetric about the resonant frequency and can be fitted well with the Lorentzian functions though fluctuations due to small signal-to-noise ratio are noticeable. As the input power increases, the resonance curve broadens gradually and becomes asymmetric and non-Lorentzian, with the peak resonant frequency shifting to lower frequency and the insertion loss increasing. Strictly speaking, when the resonance curve is clearly non-Lorentzian, Eqs. (1) and (2) become invalid and thus the traditional 3-dB bandwidth measurement of $Q_L$ is no longer applicable. In the literature, however, the 3-dB bandwidth measurement has been widely used for investigating nonlinear responses of superconducting materials without justification.

In order to extract $Q_L$ from the non-Lorentzian resonance curve, the origin of the non-Lorentzian resonance has to be examined. Firstly, due to the $T(f)$ response of the resonator, the microwave power ($P$) coupled into the microstrip varies at different frequencies around the resonance. Secondly, when $P$ is high enough, the surface impedance of the superconducting microstrip shows a power dependent behavior, namely, $Z_S=Z_S(P)$. Thus for nonlinear response, $Z_S$ is dependent on frequency. And this variation of $R_S$ and $X_S$ will cause the overall resistance $R$ and inductance $L$ of the resonant circuit to vary with frequency. As a result, the different points on the resonance curve correspond to different $LCR$ resonant circuits and thus have different effective values of $Q_L$ and $f_0$. By rewriting Eq.(1), $Q_L$ can be generally calculated as

$$Q_L = \sqrt{1/\tau - 1} f_0 / (f_R - f_L) \qquad (3)$$

where the relative power transmission ratio $\tau = T(f_R)/T(f_0) = T(f_L)/T(f_0)$ and $f_R$-$f_L$ is the bandwidth measured at $\tau$. If choosing the value of $\tau$ to be 0.5, $f_R$-$f_L$=$\Delta f_{3dB}$ and Eq.(3) turns back to Eq.(2). From the analysis above, we know that the invalidity of Eq.(3) comes from the fact that $P(f_R)$ and $P(f_L)$ are different from $P(f_0)$. It is easily to be realized that if we choose a larger value of $\tau$, smaller differences between them can be expected and Eq.(3) will give an better approximation for the $Q_L(f_0)$ value.

Shown in Fig.2 is a typical non-Lorentzian transmission curve where P indicates the resonant peak, A and A' are the 3-dB points used in the traditional $Q_L$ measurement, BB', CC' and so on indicate other pairs of reference points with different $\tau$ values. From every pair of these points, an approximate $Q_L(f_0)$ value can be obtained by measuring the bandwidth of the reference points. The accuracy of the obtained $Q_L(f_0)$ will be improved when the $\tau$ value gets larger. Theoretically, when $\tau$ of the chosen reference points is approaching 1, Eq.(3) will give an accurate $Q_L(f_0)$ value. Following the error analysis similar to that in [3], however, the relative uncertainty in measuring $Q_L$, which is the same as that in measuring the bandwidth of reference points, reads

$$\left|\frac{\Delta Q_L}{Q_L}\right| = \left|\frac{\Delta BW}{BW}\right| = \frac{1}{2\tau(1-\tau)}\Delta\tau. \qquad (4)$$

As $\tau$ is approaching 1, the error in $\tau$, which is mainly caused by the inaccuracy of the amplitude reading of the instrument, will cause very large error in the resulting $Q_L$. Here we use an extrapolation method to overcome the problem. First we measure a set of $Q_L$ values as a function of $\tau$ and then extrapolate them to a common intercept at $\tau$ =1. The resulting value is a reasonable approximation for $Q_L(f_0)$.

## 3. Results

Depicted in Fig.3 is the calculated $Q_L$ versus the chosen $\tau$ value of the reference points for the curve shown in Fig.2. It is found that when the chosen value of $\tau$ gets large, the resulting $Q_L$ value gets smaller. This agrees well with theoretical expectation for the resonator will have a larger effective $R$ at the frequency point with larger $\tau$ value. The difference between the $Q_L$ value from traditional 3-dB bandwidth measurement ($Q_L^{3dB}$) and that obtained from the method presented above ($Q_L^{ex}$) is very prominent. While the $Q_L$ value from traditional 3-dB bandwidth measurement ($Q_L^{3dB}$) is about 3200, the value obtained from the method presented ($Q_L^{ex}$) is just about 2400. The relative difference ($Q_L^{3dB} - Q_L^{ex}$)/ $Q_L^{ex}$ is larger than 30%.

We have measured the $Q_L^{3dB}$ and $Q_L^{ex}$ for the resonance curves in Fig.1. The value of ($Q_L^{3dB} - Q_L^{ex}$)/ $Q_L^{ex}$ are plotted as a function of the input power in Fig. 4 and the absolute values of $Q_L^{3dB}$ and $Q_L^{ex}$ are shown in the insert. When the power is not large, the $Q_L^{3dB}$ is just small different from $Q_L^{ex}$ and gives quite a good approximation for the $Q_L(f_0)$. As the input power increases, the difference between them gets large and $Q_L^{3dB}$ diverges noticeably from $Q_L^{ex}$. In this case, the application of the 3-dB measurement should be avoided or it will underestimate the power dependence of $Z_S$ and may even cause a misunderstanding of the origin of the nonlinearities.

## 4. Conclusions

In conclusion, we have proposed a method to extract the $Q_L$ value at the resonant peak of a non-Lorentzian resonance curve that is widely observed in the investigation of microwave nonlinear response of superconducting materials. It is a combination of a multi-bandwidth measurement and an extrapolation procedure. By analyzing the nonlinear response of a YBCO microstrip resonator as an example, it has been shown that the method proposed in this letter can do $Q_L$ extraction with dramatically improved accuracy compared with the traditional 3-dB bandwidth measurement which may cause large error in the $Q_L$ extraction when nonlinear responses are involved.

**References**

1. Ginzton, E. L. : `Microwave measurements' (McGraw-Hill, New York), 1957

2. Chin, C. C., Oates, D. E., Dresselhaus, G., and Dresselhaus, M. S. : `Nonlinear electrodynamics of superconducting NbN and Nb thin films at microwave frequencies'. Physical Reviews B, 1992, **45**, (9), pp. 4788-4798

3. Kajfez, D., Chebolu, S., Abdul-Gaffoor, M. R., and Kishk, A. A. : `Uncertainty analysis of the transmission-type measurement of Q-factor'. IEEE Transactions on Microwave Theory and Techniques, 1999, **47**, (3), pp. 367-371

**Figure Captions**

Figure 1. YBCO microstrip resonator. Transmission loss versus frequency for input power levels ranging from –18dBm to 12dBm in 5dB increments.

Figure 2. The typical non-Lorentzian resonance curve.

Figure 3. The $Q_L$ values measured as a function of $\tau$ of reference points for the resonance curve shown in figure 2.

Figure 4. $(Q_L^{3dB} - Q_L^{ex})/Q_L^{ex}$ versus input power obtained from the resonance curves in figure 1. The corresponding $Q_L^{3dB}$ (circle) and $Q_L^{ex}$ (triangle) are shown in the insert.

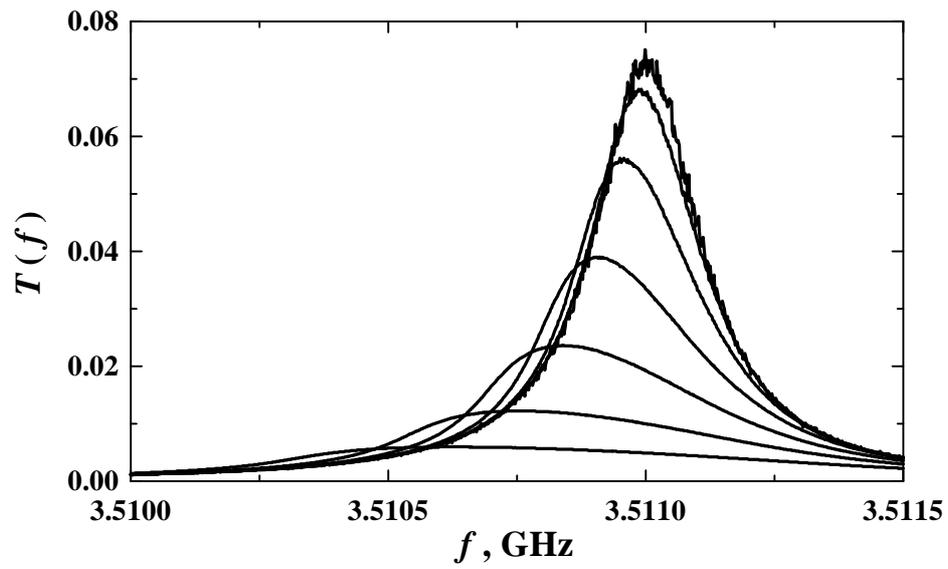

**Figure 1** Rao *et al.*

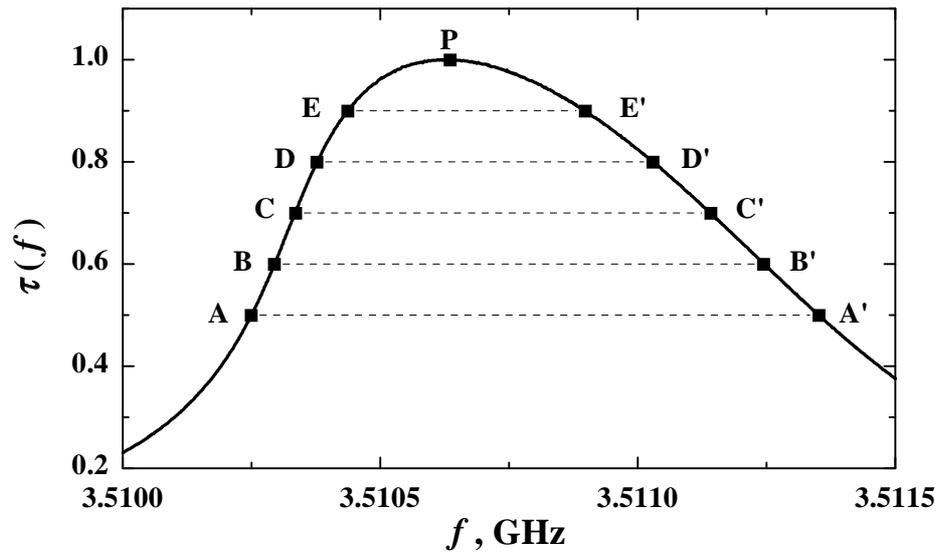

**Figure 2** Rao *et al.*

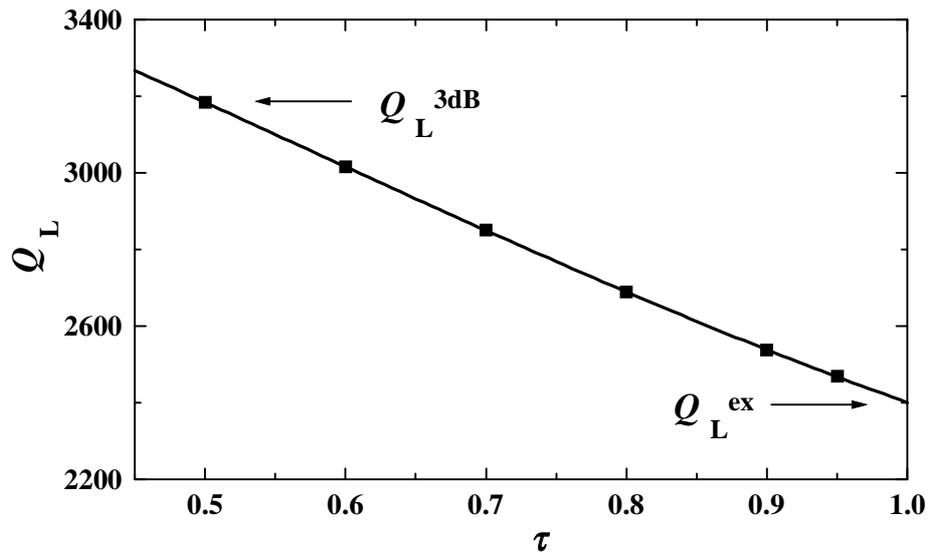

**Figure 3**    Rao *et al.*

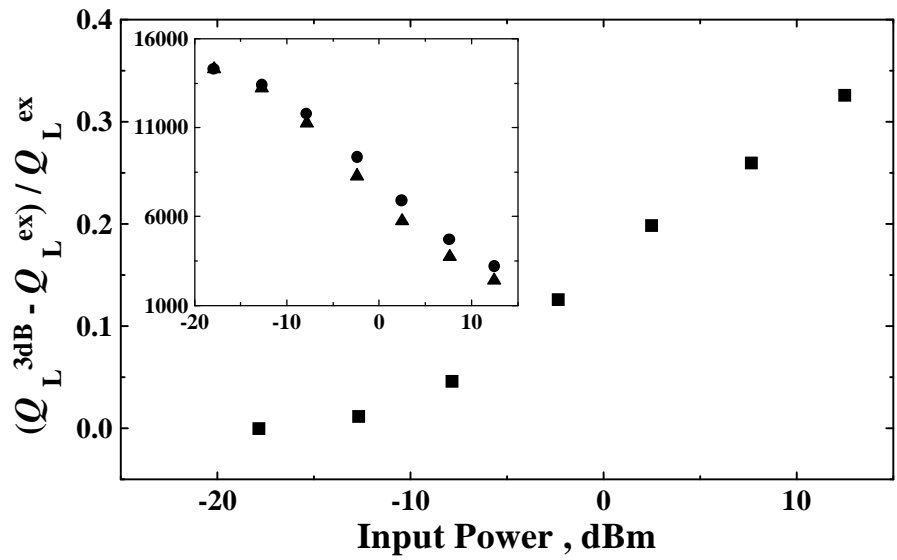

**Figure 4**     Rao *et al*.